\documentclass[12pt]{article}
\usepackage{graphicx}
\usepackage[numbers]{natbib}
\usepackage{sidecap}
\usepackage{wasysym}
\usepackage{hyperref}
\newcommand{\um}{$\mu$m}
\newcommand{\mic}{$\mu$m}
\newcommand{\apj}{ApJ}
\newcommand{\apjl}{ApJL}
\newcommand{\apjs}{ApJS}
\newcommand{\nat}{Nature}
\newcommand{\mnras}{MNRAS}
\newcommand{\physrep}{Phys. Rep.}
\newcommand{\araa}{Ann. Rev. Astron. Astropys.}
\newcommand{\pasa}{Pub. Astron. Soc. Aust.}
\newcommand{\procspie}{Proc. SPIE}

\usepackage{times}
\usepackage{geometry}
\usepackage[utf8]{inputenc}
\usepackage{enumitem,amssymb}
\usepackage{ragged2e}
\newlist{thematic}{itemize}{8}
\setlist[thematic]{label=$\square$}
\usepackage{pifont}

\begin{document}
\raggedright
\huge
Astro2020 Science White Paper \linebreak

%Understanding the nature of populations behind source-subtracted cosmic infrared background \linebreak
Populations behind the source-subtracted cosmic infrared background anisotropies \linebreak
\normalsize

\noindent \textbf{Thematic Areas:} \hspace*{60pt} $\square$ Planetary Systems \hspace*{10pt} $\square$ Star and Planet Formation \hspace*{20pt}\linebreak
$\square$ Formation and Evolution of Compact Objects \hspace*{31pt} $\XBox$ Cosmology and Fundamental Physics \linebreak
  $\square$  Stars and Stellar Evolution \hspace*{1pt} $\square$ Resolved Stellar Populations and their Environments \hspace*{40pt} \linebreak
  $\XBox$    Galaxy Evolution   \hspace*{45pt} $\square$             Multi-Messenger Astronomy and Astrophysics \hspace*{65pt} \linebreak
  
\textbf{Principal Author:}

Name:	A. Kashlinsky
 \linebreak						
Institution:  Code 665, Observational Cosmology Lab, NASA Goddard Space Flight Center, Greenbelt, MD 20771 and SSAI, Lanham, MD 20770
 \linebreak
Email: Alexander.Kashlinsky@nasa.gov
 \linebreak
Phone:  301-286-2176
 \linebreak
 
\textbf{Co-authors:} %(names and institutions) 
R. G. Arendt (GSFC and UMBC), 
M. Ashby (CfA), 
F. Atrio-Barandela (U. Salamanca), 
V. Bromm (UT Austin), 
N. Cappelluti (U. Miami), 
S. Clesse (UCLouvain), 
A. Comastri (INAF), 
J-G. Cuby (CNRS),
S. Driver ( UWA),
G. Fazio (CfA),
A. Ferrara (Pisa), 
A. Finoguenov (U. Helsinki), 
D. Fixsen (GSFC and UMCP), 
J. Garcia-Bellido (IFT-UAM/CSIC),
G. Hasinger (ESA), 
K. Helgason (U. Iceland), 
R. J. Hill (GSFC),  
%K. Jahoda, 
R. Jansen (ASU), 
%E. Komatsu (MPA), 
J. Kruk (GSFC), 
%M. Markevich (GSFC), 
J. Mather (GSFC), 
%T. Matsumoto, A. Merloni (MPE), S.H. Moseley (GSFC), 
P. Natarajan (Yale), 
N. Odegard (GSFC and ADNET), 
%R. Petre (GSFC), 
T. Reiprich (U. Bonn), 
M. Ricotti (UMCP), 
M. Sahlen (U. Uppsala),
E. Switzer (GSFC),
%M. Urry (Yale), 
R. Windhorst (ASU), 
%J. Wise,
 E. Wollack (GSFC), 
 Bin Yue (NAOC)
\linebreak

\justifying
%\textbf{Abstract  (optional):}
\noindent
\textbf{Abstract:}
Although the advent of ever-larger and more sensitive telescopes in the coming decade will reveal
correspondingly fainter, more distant galaxies, a question will persist: what more is there that
these telescopes cannot see?  One answer is the source-subtracted Cosmic 
Infrared Background (CIB).  The CIB is comprised of the collective light from all sources remaining 
after known, resolved sources are accounted for.  A crucial point: unlike the cosmic microwave 
background, the CIB arises from discrete sources.  Ever-more-sensitive surveys will identify the brightest
of these, allowing them to be removed, and -- like peeling layers off an onion -- reveal deeper
layers of the CIB.  In this way it is possible to measure the contributions from populations not
accessible to direct telescopic observation.  Measurement of fluctuations in the source-subtracted CIB, 
i.e., the spatial power spectrum of the CIB after subtracting resolved sources, provides a robust 
means of characterizing its faint, and potentially new, populations.  Studies over the past 15 years 
have revealed source-subtracted CIB fluctuations on scales out to $\sim100'$ which cannot be explained by extrapolating from 
known galaxy populations.  Moreover, they appear highly coherent with the unresolved Cosmic X-ray 
Background, hinting at a significant population of accreting black holes among the CIB sources.
{\bf Characterizing the source-subtracted CIB with high accuracy, and thereby constraining
the nature of the new populations, is feasible with upcoming instruments and would produce 
critically important cosmological information in the next decade.} New coextensive 
deep and wide-area near-infrared, X-ray, and microwave surveys will bring decisive
opportunities to examine, with high fidelity, the spatial spectrum and origin of the CIB 
fluctuations and their cross-correlations with cosmic microwave and X-ray backgrounds, 
and determine the formation epochs and the nature of the new sources 
(stellar nucleosynthetic or accreting black holes).

\pagebreak
\clearpage
\centerline{\bf 1. Introduction}

The cosmic infrared background (CIB) includes emissions from objects inaccessible to direct telescopic studies \cite[see review by][and refs therein]{Kashlinsky:2005}. However, direct measurements of the CIB intensity provide only upper limits 
at most near- to mid-IR wavelengths 
because of uncertainties in the contributions of Galactic and zodiacal 
foregrounds. A complementary approach is to characterize the spatial fluctuations of the 
source-subtracted CIB \cite[][]{Kashlinsky:1996a}. This analysis can employ data sets without an
accurate determination of the absolute zero point, and avoids some of the
difficulties in modeling the foreground contributions \cite[][]{Kashlinsky:1996,Kashlinsky:2000,Hauser:1998,Matsumoto:2005,Arendt:2016}. The spatial power 
spectrum of CIB fluctuations depends on the 
clustering of the remaining sources, and their 
integrated emission. 

\begin{SCfigure}[1.0][b!]
%\centering
\hspace{-1.cm}
\includegraphics[width=4.75in,height=2.1in]{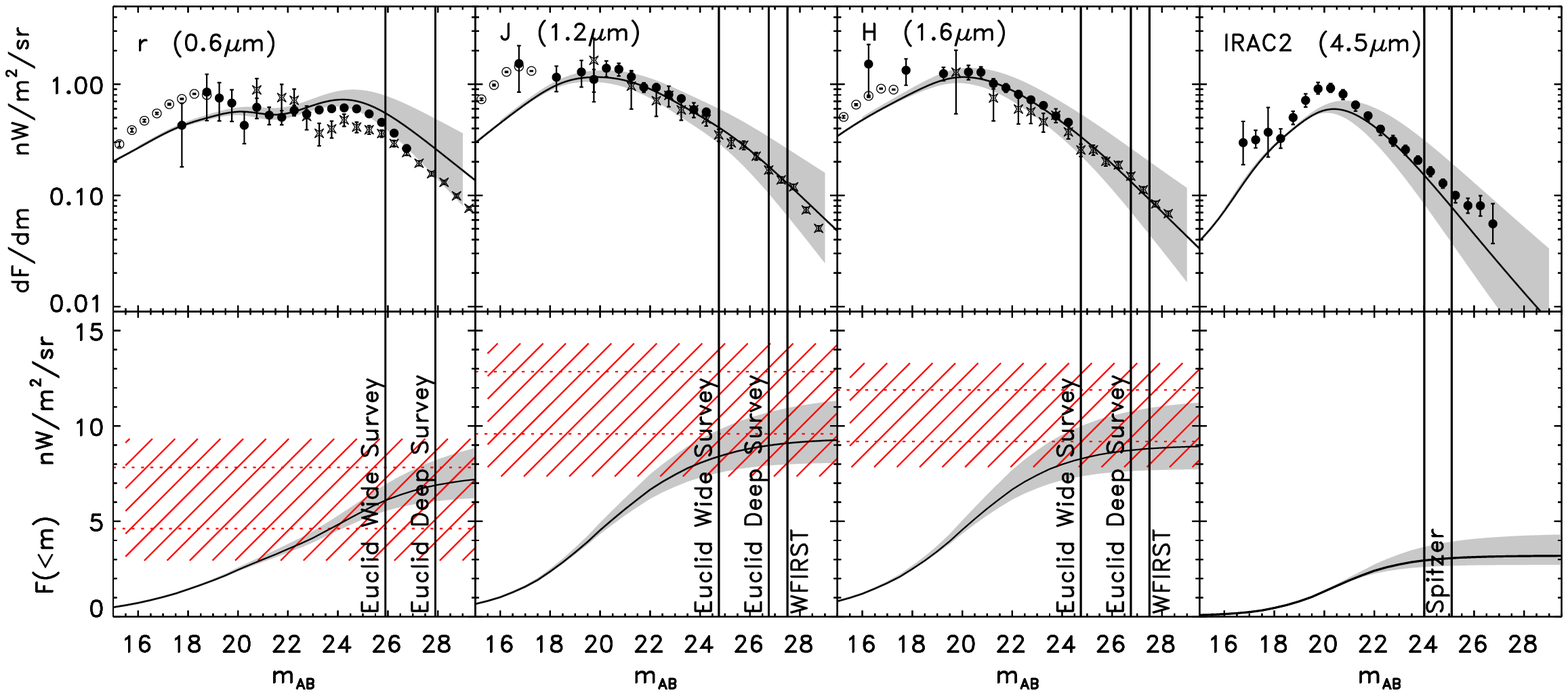}
\vspace{-0.25cm}
\caption{\scriptsize 
{\bf Top}: Data points show the differential contribution to the mean background flux for
galaxy count surveys \cite{Driver:2016}. The line and shaded uncertainty band 
are the mean flux provided by the reconstruction of \cite{Helgason:2012a}.
{\bf Bottom}: The reconstruction  in black shows the net  
EBL flux vs. magnitude. The red dotted lines (hatched band) show the 1$\sigma$
(2$\sigma$) range on EBL estimates from $\gamma$-ray opacity
\citep{Fermi-LAT-Collaboration:2018} (the 4.5\mic\ limits are not constraining and are not shown). The range between the lowest limit 
from the integrated counts and the highest limit of the $\gamma$ rays 
is the possible intensity of the source-subtracted extragalactic light.
\label{fig:cib_counts}
}
\end{SCfigure}
Lower limits on the extragalactic background (EBL)  flux come from integrating 
source counts in various bands; upper bounds can be estimated from 
the $\gamma$-ray opacity caused by the CIB/EBL (Fig. 1). 
All these estimates are generally consistent, and the present 
uncertainties show the range of the allowed CIB contribution from sources fainter
than detected in surveys. The source-subtracted CIB is that remaining 
after subtraction of the contribution from individually resolved sources, and 
is the key observable information on an otherwise unseen portion of the Universe.

The efforts over the past decade and a half identified source-subtracted CIB fluctuations in deep {\it Spitzer} and {\it Akari} data from {\it new} unknown populations. The measurements \cite[][]{Kashlinsky:2005a,Kashlinsky:2007a,Kashlinsky:2007,Kashlinsky:2012,Arendt:2010,Matsumoto:2011,Cooray:2012} span 2--5\mic; at shorter wavelengths there is currently significant uncertainty with conflicting results from, in chronological order, deep 2MASS, {\it HST}/NICMOS, CIBER and {\it HST}/WFC3 analyses \cite[][]{Kashlinsky:2002,Odenwald:2003,Thompson:2007,Thompson:2007a,Zemcov:2014,Mitchell-Wynne:2015}. The source-subtracted CIB from deep {\it Spitzer} data appears highly coherent with soft cosmic X-ray background (CXB) \citep{Cappelluti:2013,Cappelluti:2017,Mitchell-Wynne:2016,Li:2018} implying a significant presence of accreting black holes (BHs) among the new CIB sources. (See review by \cite{Kashlinsky:2018}.)

The CIB fluctuation signal implies new populations below the (faint) flux limit (AB$>\atop\sim$25) with significant implications for cosmology. {\it Identifying with high accuracy the properties of source-subtracted CIB and understanding the nature of its populations is feasible with upcoming instruments and should be one of the major goals in cosmology for the coming decade.}

Theoretically such CIB signal was predicted to arise from the first stars era (FSE) \cite{Kashlinsky:2004,Cooray:2004}. The sources responsible for it can come from Population III, predicted to be very massive stars  that, for the standard $\Lambda$CDM model, form in first collapsed minihalos of $10^{6-9}M_\odot$ at $z> 10$ \cite{Bromm:2004,Bromm:2001,Abel:2002}; the faint minihalos will have high projected surface density lying largely in the confusion noise of the next decade telescopes \cite{Kashlinsky:2015a}. The early epochs could also contain abundant BHs of various origins, contributing at both IR and X-ray \cite{Agarwal:2012,Yue:2013,Yue:2014,Yue:2016a,Kashlinsky:2016,Latif:2016}. Also there may be contributions from new stellar populations at low to intermediate $z$ as well as from a new particle decay \cite{Bond:1984,Cooray:2012,Gong:2015}. The populations cannot be observed directly by existing telescopes, but could be probed via source-subtracted CIB anisotropies (or fluctuations).

\centerline{\bf 2. Properties of source-subtracted CIB anisotropies and cosmological implications}

\begin{figure}[t]
%\plotone{fig_shotnoise.eps}
%\includegraphics[width=4in]{fig-vs_sn.eps}
%\hspace{-10mm}\includegraphics[width=4.7in]{fig2_sn.eps}\hspace*{4mm}
\hspace{-10mm}\includegraphics[width=4.7in]{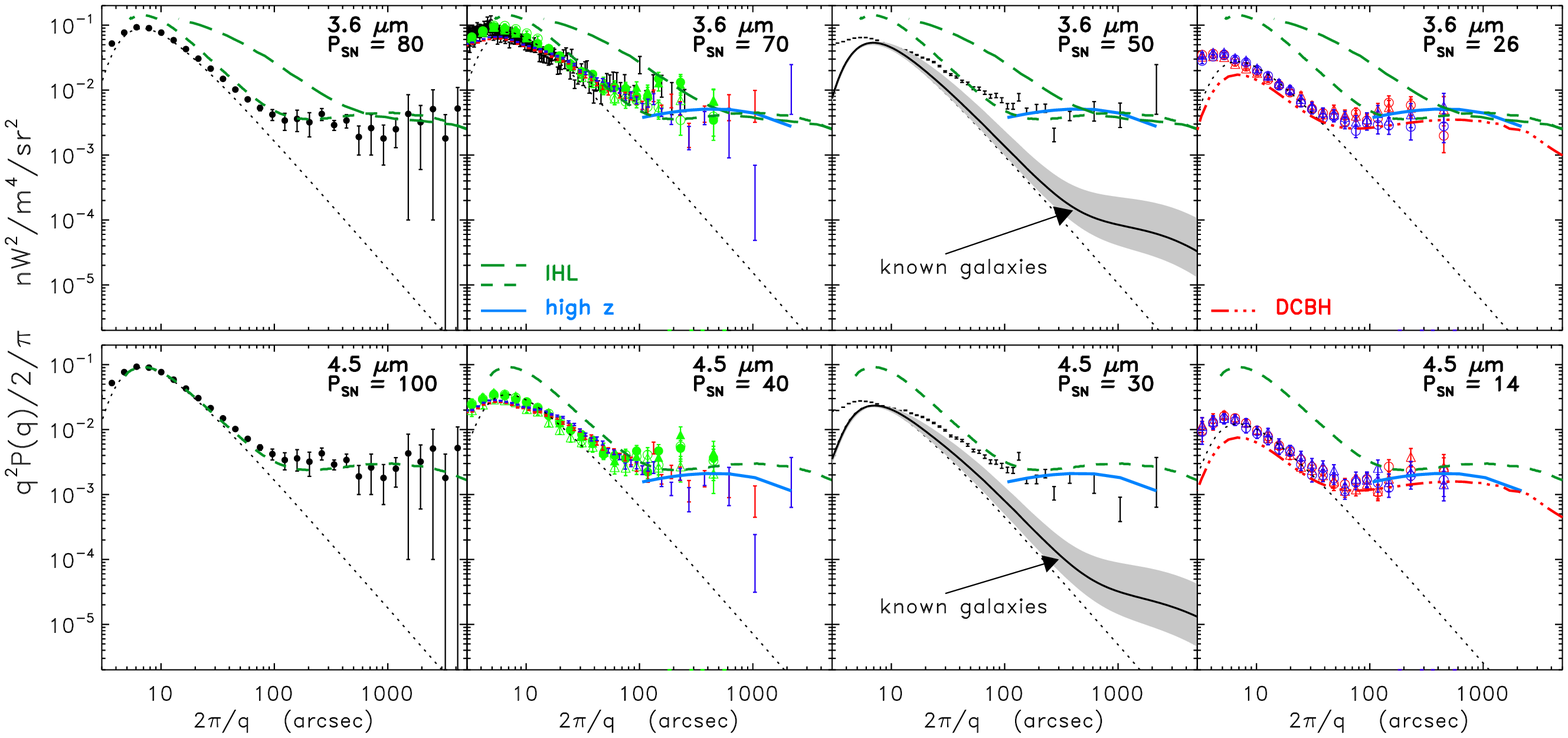}\hspace*{4mm}
\includegraphics[width=2in,height=2.125in]{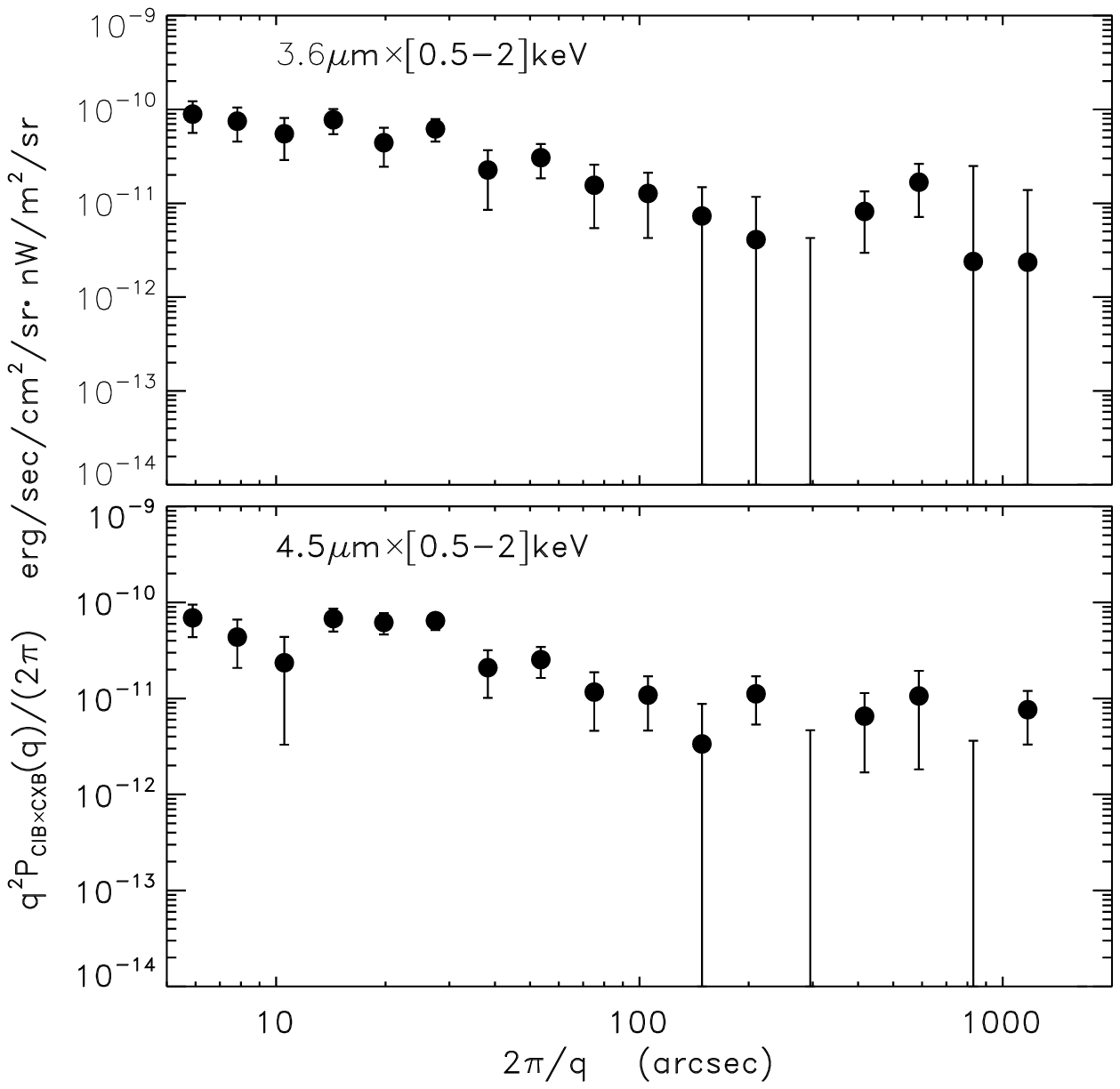}
\caption{\scriptsize Mean squared fluctuation, $q^2P(q)/(2\pi)$, at angular scale $2\pi/q$. {\bf Left}: {\it Spitzer}-based source-subtracted CIB fluctuations at different shot-noise power levels \cite{Kashlinsky:2005a,Kashlinsky:2007a,Kashlinsky:2012,Cooray:2012} at 3.6\um\ and 4.5 \mic.  {\it No decrease of the large-scale clustering component with shot noise yet appears at these shot-noise levels}, its power marked in nJy$\cdot$nW/m$^2$/sr. Contribution from remaining known galaxies is shown from \cite{Helgason:2012a}.
%Black solid lines show the shot noise component from remaining known galaxies at $m_{\rm AB}\geq 28$ at the two longest NIRCam wavelengths that will reached the 
%configuration proposed in Sec.\ \ref{sec:jwst_pars}. 
The intrahalo light (IHL) models at low $z$, green short dashes from \cite{Cooray:2012a}  and green long dashes \cite{Zemcov:2014} , in their presented form 
appear inconsistent with the CIB data and cannot account for the CIB-CXB cross-power shown in right panels. 
 BH-based models can account for the data at all shot-noise levels and the observed CIB-CXB cross-power. The DCBH model \citep{Yue:2013} is shown with red dash-triple-dotted lines, the LIGO-type PBH model \citep{Kashlinsky:2016} is shown with blue solid line. Adapted from \citep{Kashlinsky:2018}.
 {\bf Right}: The CIB-CXB cross-power between the IRAC 3.6 and 4.5 \mic\ CIB and Chandra soft CXB \citep{Cappelluti:2017}.
}
\label{fig:cib_irac}
\end{figure}

The source-subtracted CIB fluctuations uncovered in deep {\it Spitzer} data, are displayed in Fig. \ref{fig:cib_irac}, left. They contain two terms:  1) shot-noise, or the convolved white noise, and 2) the clustering term which originates from new unknown populations. The {\it Spitzer}-  and {\it Akari}-based measurements \cite[][]{Kashlinsky:2005a,Kashlinsky:2007a,Kashlinsky:2012,Arendt:2010,Matsumoto:2011,Cooray:2012} cover 2--5 \mic;
%; at shorter wavelengths there is currently significant uncertainty with conflicting results from, chronologically ordered, deep 2MASS, HST/NICMOS, CIBER and HST/WFC3 analyses \cite[][]{Kashlinsky:2002,Odenwald:2003,Thompson:2007,Thompson:2007a,Zemcov:2014,Mitchell-Wynne:2015}. 
their main established properties are summed up below with Fig. \ref{fig:cib_irac} showing their most relevant properties \cite[see ref. ][Sec. V.B.3]{Kashlinsky:2018}:
\\
$\bullet$ The shot noise component, $P_{\rm SN}$=$\int_{m_0}^\infty S^2(m)dN$, dominates small angular scales (dotted lines in Fig.\ref{fig:cib_irac}, left panels). It still comes mainly from the  known sources below the limiting flux $S$ of magnitude $m_0\!\simeq$24--25  (related to survey flux limits).\\ %The larger scales reflect the clustering component of the new sources.
$\bullet$ There is a clear excess of CIB clustering power (scales $\gtrsim 100''$) at 3.6 and 4.5 \mic\ over that from known remaining galaxies \citep{Kashlinsky:2005,Helgason:2012a}. The excess power from clustering appears isotropic on the sky consistent with a cosmological origin \citep{Kashlinsky:2007a,Kashlinsky:2012}. 
%Small-scale power ($\lesssim 100''$) is well accounted for by the shot-noise (related to survey flux limits) from the remaining known galaxies.
\\
$\bullet$ The clustering component does not yet appear to drop with decreasing shot-noise suggesting it is produced by very faint sources with 3.6 and 4.5 \mic\ flux densities $\lesssim$20nJy \citep{Kashlinsky:2007b}. %See Fig. \ref{fig:cib_irac}, left.
\\
$\bullet$ The CIB fluctuations at {\it AKARI} wavelengths suggest a Rayleigh-Jeans energy spectrum for the power at 2--5\mic, i.e. $P\!\propto\!\lambda^{-2n}$ with $n\!\sim$3  \citep{Matsumoto:2011}. \\
$\bullet$ {\it Spitzer/AKARI} band-integrated CIB fluctuations are $\delta F_{2-5\mu{\rm m}}(5')\simeq$0.1 nW/m$^2$/sr \citep{Kashlinsky:2018}, corresponding to the mean CIB flux $\sim$1 nW/m$^2$/sr \citep{Kashlinsky:2007b} if at high $z$, which added to known populations is consistent with the $\gamma$-ray absorption measurements of $11.6^{+2.6}_{-3.1}$ nW/m$^2$/sr (2$\sigma$) at 1.4 \mic\ \citep{Fermi-LAT-Collaboration:2018}. \\
$\bullet$ The CIB clustering component is strongly coherent with the (soft) cosmic X-ray background (CXB) \citep{Cappelluti:2013,Cappelluti:2017,Mitchell-Wynne:2016,Li:2018}. The cross-power, $P_{\rm CIB\times CXB}$, cannot be explained by remaining known galaxies \citep{Helgason:2014}. The emerging CXB-CIB coherence, ${\cal C}\!\equiv\!\!P_{\rm CIB\times CXB}^2/P_{\rm CIB}/P_{\rm CXB}$, exceeds ${\cal C}\!\gtrsim$0.04 \citep{Cappelluti:2017} suggesting a significant abundance of accreting BHs among the CIB-producing sources. 
\\
\centerline{\bf 3. Questions posed by the CIB fluctuation measurements}
The current theoretical proposals for the origin of the sources behind the CIB fluctuations differ in the epochs populated by these sources: the original discovery by \cite{Kashlinsky:2005} posited that the signal is from the first stars era, while later \cite{Cooray:2012} proposed that it comes from an intrahalo light (IHL) of normal stars stripped away in galactic mergers at low to intermediate $z$. The IHL proposals (Fig. \ref{fig:cib_irac}), as presented, do not fit the CIB fluctuations at lower shot noise levels, do not account for the CIB-CXB coherence, and the modelling in \cite{Zemcov:2014} appears in tension with the the newest $\gamma$-ray absorption limits \citep{Fermi-LAT-Collaboration:2018}. Massive Pop III stars at $z\gtrsim$10 can account for the 
CIB fluctuations 
%signal, in the absence of BH contributions, 
only with somewhat ``optimistic" efficiencies of formation (5-10\%) inside the first minihalos \citep{Helgason:2016}. With the discovery of the CIB-CXB coherence, two suggestions have been made for the origin of these fluctuations, both involving BHs at high $z$: 1) direct collapse BHs (DCBHs) \citep[][]{Yue:2013} and 2) LIGO-type primordial BHs (PBHs) making up dark matter \citep{Kashlinsky:2016}. 
%The subject is thus important to proper
CIB fluctuations are thus a critical tool for the understanding of: 
i) new populations that cannot be detected directly; 
ii) the physics of the pregalactic Universe; 
iii) the nature of the first sources and, potentially, reionization; 
iv) BH activity; 
and v) the possible connection to the nature of dark matter and LIGO-type BHs.

The specific questions that need to be answered, and the required configurations, are:\\
\centerline{{\bf Q1:} {\it What are the epochs of the sources producing the CIB fluctuations?}}
The redshift of the sources producing the CIB fluctuations can be estimated by identifying
a Lyman break in the spectrum of the sources, in much the same way that individual galaxy
redshifts can be estimated from photometric dropouts. However, to be effective for the 
CIB, there must be a relatively sharp cutoff for the minimum $z$ of the sources,
and the foreground sources that would not show a Lyman break must be minimized. Observations 
in the visible and near-IR are needed, as sources whose epoch ends at $z\sim10$ would show a Lyman break at $\sim 1$ \mic.
The observational configuration required to isolate the Lyman break must reach AB$\gtrsim$25 since at brighter magnitudes the CIB power below $\sim$1\mic\ from remaining known galaxies, and its systematic uncertainty strongly dominates that observed from the new populations at 2--5\mic\ \citep{Kashlinsky:2015a}.\\
\centerline{{\bf Q2:} {\it How does the CIB from these sources evolve over cosmic time?}}
The redshift evolution of the CIB fluctuations can be isolated in further detail  via Lyman tomography by 
examining, suitably subtracting, the relative brightness of the large-scale fluctuations in a series of adjacent
spectral bands.%, the subsequent wavelength then probing additional emissions from an extra range of $z$. 
With increasing wavelength, such comparisons are sensitive to the 
fraction of the population lying at increasingly high redshifts
\citep{Kashlinsky:2015a,Kashlinsky:2015}. These studies require deep imaging 
(to remove foreground sources) over large areas (to obtain high $S/N$ power 
spectra at large scales) in many adjacent filters (to explore as a function of $z$).\\
\centerline{{\bf Q3:} {\it What are the relative contributions of stars and BHs to the CIB fluctuations?}}
The CIB-CXB cross-correlation appears real, but not sufficiently constrained at present. 
Improved measurements of the correlations at large and small angular scales can better
resolve the CIB fraction produced by sources that are associated 
with the X-ray emission. Whether this correlation is found only in the large-scale 
clustering component or also extends to small angular scales, due to intrinsic coherence of the CIB and CXB shot noise from the new sources, will indicate 
the extent to which the IR and X-ray sources are physically the same 
objects or whether they are distinct objects (emission mechanisms) that are 
grouped together through the cosmic structure
\citep{Helgason:2016,Yue:2013,Kashlinsky:2015,Windhorst:2018}.\\
\centerline{{\bf Q4:} {\it What is the contribution of the new sources to the CXB fluctuations?}}
In the event that the intrinsic CIB-CXB coherence is high and the CIB fluctuations are 
measured with better $S/N$ than the CXB fluctuations, the CIB fluctuations can
be used to make a better estimate of the CXB fluctuations from the correlated
sources than is possible from the X-ray observations alone. See Figure \ref{fig:cxb/cmb},left.\\
\centerline{{\bf Q5:} {\it What is the contribution of the new sources to reheating and 
reionization of the IGM?} }
The IGM plasma would produce weak temperature anisotropies in the cosmic microwave background (CMB) via the Sunyaev-Zeldovich (SZ) effect. While these SZ anisotropies are too faint to be detected,  the cross-correlation of maps of specific source-subtracted CIB fluctuations with suitably constructed microwave maps at different frequencies, can probe the physical state of the gas during reionization and test/constrain models of the early CIB sources \citep{Atrio-Barandela:2014}. \\
\centerline{{\bf Q6:} {\it Can the CIB fluctuations constrain the cosmological model at high $z$?}}
Lyman tomography uses measurements of the CIB fluctuations over large areas at multiple wavelengths to derive high-precision measurements of the CIB power spectrum as a function of redshift at $z\!>$10.
This in turn may enable probing the baryonic acoustic oscillations (BAOs) at redshifts beyond those that can be probed by other means, opening a brand new window on the standard cosmological model to obtain constraints from the high redshift universe \citep{Kashlinsky:2015}.\\
\centerline{\bf 4. Outlook to the coming decade}

\begin{figure}[b!]
%\plotone{fig_shotnoise.eps}
\hspace{-0.35cm}
\includegraphics[width=6.5in]{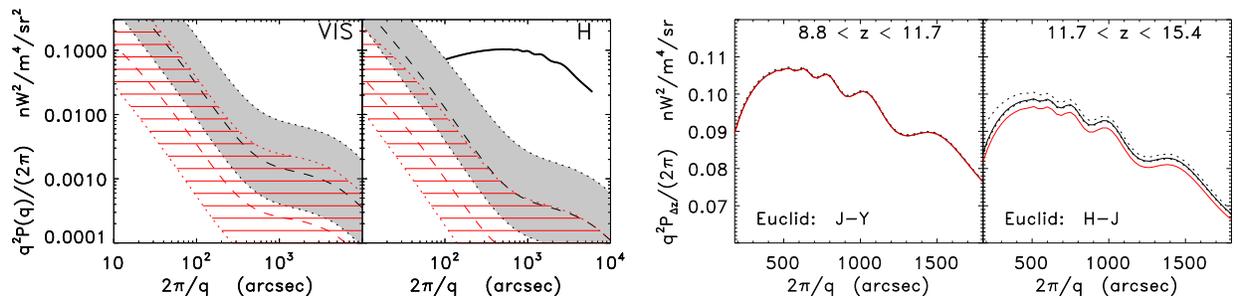}
\caption{\scriptsize  {\bf Left}: CIB fluctuations from known galaxies remaining 
in the {\it Euclid} VIS and NISP bands (grey shaded area for Wide Survey and red lined area for Deep Survey). Thick solid line is a high-$z$ CIB, which fits {\it Spitzer} 3.6, 4.5 \mic\ CIB fluctuations. {\bf Right}: The Lyman-tomography reconstruction of the CIB emission history and BAOs for {\it Euclid}'s (Y,J,H) filters and Wide Survey depth at each $z$-range.  Red line shows the underlying CIB fluctuations by sources in the marked $z$-range from high-$z$ stellar populations reproducing {\it Spitzer} measurements. Black lines include also the contributions from known remaining galaxies \cite{Helgason:2012a} with its uncertainty marked by dotted lines. Adapted from \citep{Kashlinsky:2018}. }
\label{fig:cib_future}
\end{figure}

In the next decade, new observations of the CIB fluctuations will need to 
meet several requirements to significantly help answer Q1 and Q2, above.
First, the observations will need to be made over large areas, on order of 
$10^3$~deg$^2$. This is needed to obtain the best possible $S/N$ out to the 
largest possible angular scales. The data must be collected and processed 
by means which accurately capture large scale structure \citep[][]{Fixsen:2000,Arendt:2000,Arendt:2002}. Second, 
observations in multiple near-IR bands are needed to study the SED of the 
fluctuations to probe the cutoff redshift (Q1), and to 
determine the cosmic history of the emission of the CIB sources.
Third, the data should have sufficient sensitivity and angular resolution 
to detect and subtract as much of the total CIB as possible. 

Fig.\ref{fig:cib_future} shows how the data from the upcoming {\it Euclid} 
mission \citep[][]{Laureijs:2011,Laureijs:2014} will provide very useful results via the NASA selected LIBRAE (Looking at Infrared Background Radiation Anisotropies with {\it Euclid}) project ({\tiny {\bf \url{https://www.euclid.caltech.edu/page/Kashlinsky%20Team}}}). 
{\it Euclid}'s NISP instrument will probe CIB fluctuations 
in Y, J, and H near-IR bands, and VIS which will do the same in 
a single broad visible light band. The left panels of Figure \ref{fig:cib_future}
show the expected contributions of remaining galaxies in the CIB power 
spectrum obtained from {\it Euclid's} Wide and Deep surveys,
after subtraction of sources to the  
AB$\simeq$25--26 magnitude limits. 
In the H band, the extrapolated large-scale fluctuation is expected 
to be much stronger than that of faint known galaxies.
In the VIS band, the clustering component will drop out if due to high $z$ sources, and only the power
of the unresolved known populations should be detected. The right panels of Fig.\ref{fig:cib_future}
show that the 3 near-IR bands will enable Lyman tomography to explore the 
history of the emission (Q2), and the large areas will allow sufficient 
$S/N$ to probe the BAO imprint in the power spectrum, addressing (Q6).

\begin{figure}[t!]
%\plotone{fig_shotnoise.eps}
%\hspace{-1.5cm}
%\includegraphics[width=3in, trim= 1.5cm 0 0 0]{fig1.eps}
%\includegraphics[width=4.5in, trim= 1.5cm 0 0 0]{fig1_v1.eps}\\
%\includegraphics[width=1.825in,trim=0 0 -100 0,clip=true]{fig2b.eps}
\includegraphics[width=4.2in]{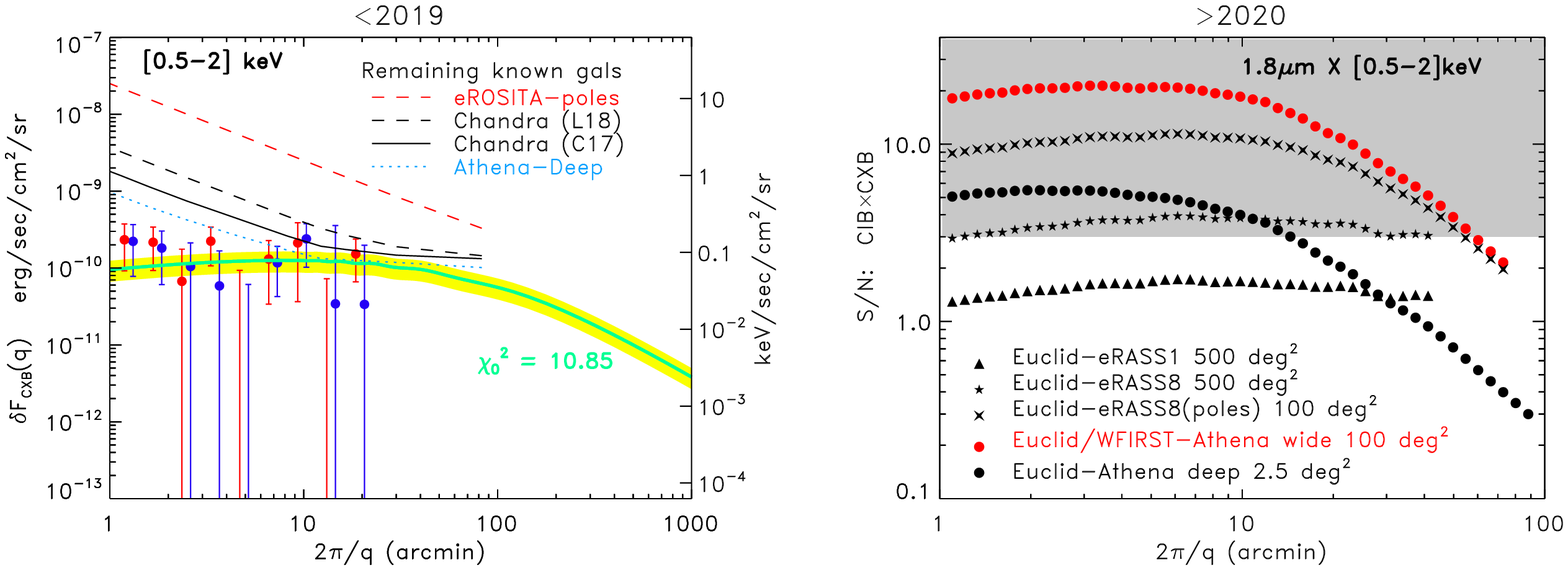}
\hspace{5mm}
\includegraphics[width=2.in]{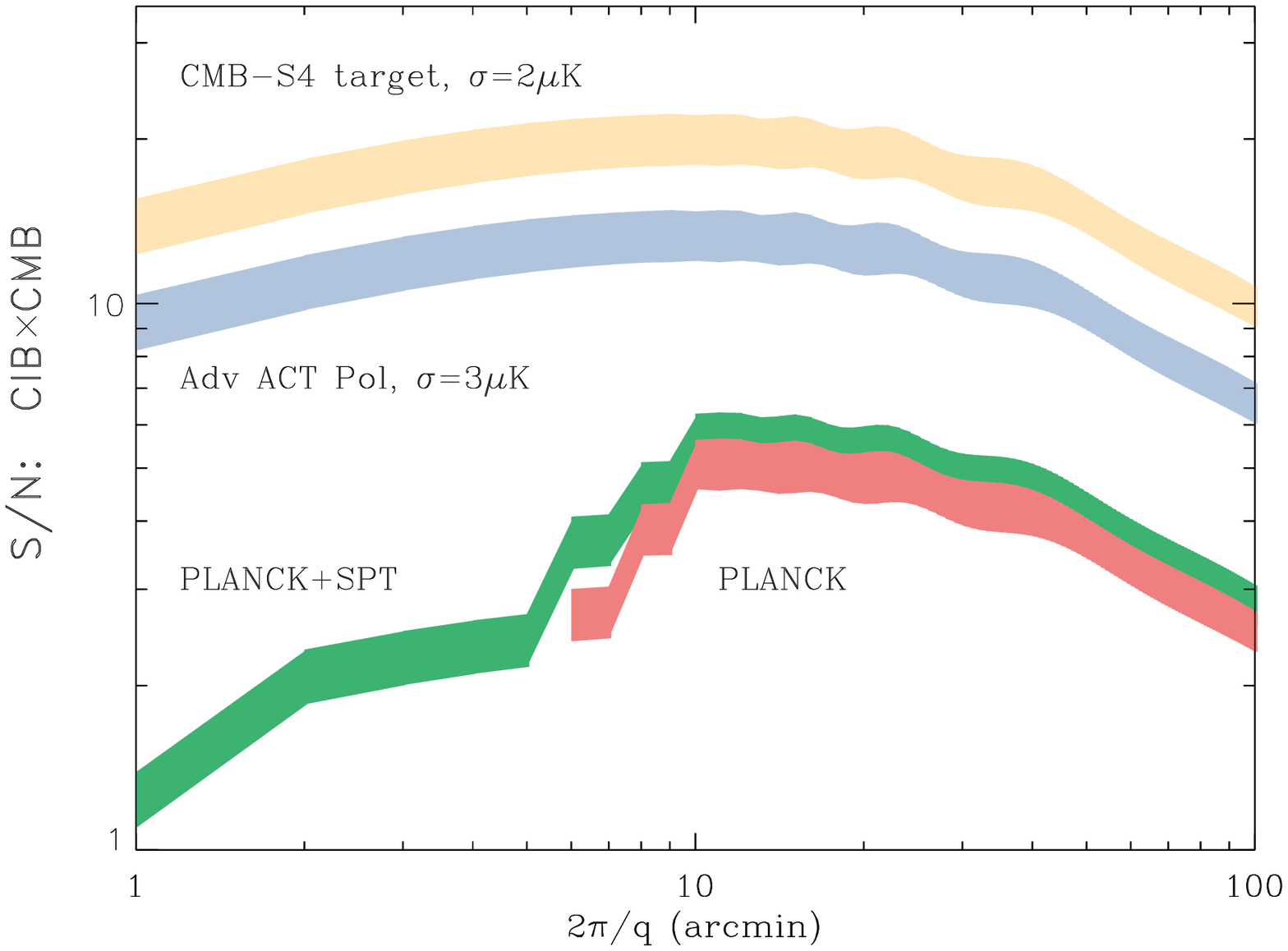}
\caption{\scriptsize  {\bf Left}: Residual CXB, $\delta F_{\rm X}\!\equiv \!(q/\sqrt{2\pi}) P_{\rm CIB\times CXB}/\sqrt{P_{\rm CIB}}$, from the new populations \citep{Kashlinsky:2019}. Blue (3.6 \mic) and red (4.5 \mic) circles are derived using the IRAC/{\it Chandra} measurements \citep{Cappelluti:2017}.  Solid green line is the best fit for the 3D $\Lambda$CDM power template  at $d_A=7$ Gpc ($z$=15); the yellow regions marks 1$\sigma$ deviation of the fit which gives $\chi^2_0=10.85$ for the 18 data points. 
The angular scales shown here are where clustering component dominates, and at the same exceeds the contributions from known sources at $>5\sigma$ level for both IR bands. The CXB fluctuations from known sources remaining in the marked configurations are shown to be above the signal if probed directly. From \cite{Kashlinsky:2019}.  {\bf Middle}: Overall S/N for the signal in the right panel through CIB-CXB cross-power measurements from {\it Euclid-eROSITA} and {\it Euclid-Athena} configurations. Shaded area marks $S/N \geq 3$. From \cite{Kashlinsky:2019}. {\bf Right}: Filled regions show the range of the S/N of the CIB-TSZ cross power over the Euclid Wide Survey region for the marked experimental configurations for an IGM temperature of $T_{\rm e}$=$10^4$K; the value of $S/N\propto T_{\rm e}$. From \cite{Atrio-Barandela:2014,Kashlinsky:2018}. }
\label{fig:cxb/cmb}
\end{figure}
Substantial improvements in the CIB-CXB cross-power will be provided by the {\it Euclid} CIB data and {\it eROSITA} \citep[][]{Merloni:2012} X-ray maps. These would bring decisive probes of the cross-power signal \citep[][]{Kashlinsky:2019} particularly from the deep {\it eROSITA} coverage of 140 deg$^2$ at the poles. ESA's {\it Athena} \citep[][]{Nandra:2013}, to be launched later in the next decade, will bring further refinements in the measurement, particularly if it covers a wide area of $\sim$100 deg$^2$ at the depth corresponding to the current {\it Chandra} integrations in \citep[][]{Cappelluti:2013,Mitchell-Wynne:2016,Cappelluti:2017,Li:2018}. Figs.\ref{fig:cxb/cmb},left/middle show the expected progress for answering Q3,Q4. The large areas covered by {\it eROSITA} and {\it Athena} with good energy resolution will allow probing the CIB-CXB cross-power at hard X-ray energies, enabling further insights into Q3,Q4 and the new sources.

If they are at high-$z$, further probe of the CIB sources will involve testing their impact on the state of IGM at reionization by measuring the thermal SZ component using a cross-correlation of the source-subtracted CIB from {\it Euclid} with multifrequency CMB maps, suitably constructed to remove primary and kinematic SZ CMB terms \citep[][]{Atrio-Barandela:2014}. The prospects here are illustrated in Fig. \ref{fig:cxb/cmb},right which shows the possibility of probing IGM temperature to well below 10$^4$K with multifrequency CMB maps of low noise and high resolution from AdvACTPol \citep[][]{Ward:2016,Thornton:2016,The-Simons-Observatory-Collaboration:2018} and CMB-S4 \citep[][]{Abitbol:2017}.

Further CIB information will be available from {\it WFIRST} to be launched in 2nd half of 2020's \citep[][]{Spergel:2015} which will map 2,000 deg$^2$ with 4 filters from 0.9 to 2 \mic. The area covered will be smaller, 
but observed more deeply than in the {\it Euclid} surveys. The {\it WFIRST} wavelength coverage will extend 
to longer wavelengths allowing the probe of potentially higher-$z$ components.
The areal coverage of {\it JWST} surveys will be too limited to provide a
  detailed probe of the clustering signal in the power spectrum.
However {\it JWST} will provide the 
deepest possible direct survey of contributors to the CIB. A configuration covering 1 deg$^2$ to $m_{\rm AB}=28$ in all 7 NIRCAM wide filters will take $\sim$400 hrs of {\it JWST}'s time and was proposed by \citep[][]{Kashlinsky:2015a} to result in significant new information (Q1,Q2) on the new CIB sources, including their energy spectrum over 0.6--6\mic. 

New ground-based telescopes can provide very deep and wide surveys
of the resolved galaxy populations, but are not well-suited for measuring CIB fluctuations at the scales and wavelengths of interest.
The study of CIB fluctuations will perfectly complement ongoing neutral hydrogen 21 cm line intensity mapping studies (e.g. PAPER, EDGES, HERA, LOFAR, MWA, SKA), where the 21 cm global or power spectrum signals characterize the topology and $z$-evolution of the IGM \citep[][and refs therein]{Mesinger:2016}. Since CIB is more sensitive to the nature and properties of early sources, the combination of the two approaches will provide a complete picture at the time of the first light. 

\pagebreak
%\textbf{References}
%\bibliography{references_wp-cib} 

\begin{thebibliography}{63}
\providecommand{\natexlab}[1]{#1}
\providecommand{\url}[1]{\texttt{#1}}
\expandafter\ifx\csname urlstyle\endcsname\relax
  \providecommand{\doi}[1]{doi: #1}\else
  \providecommand{\doi}{doi: \begingroup \urlstyle{rm}\Url}\fi

\bibitem[{Abel} et~al.(2002){Abel}, {Bryan}, and {Norman}]{Abel:2002}
T.~{Abel}, G.~L. {Bryan}, and M.~L. {Norman}.
\newblock {The Formation of the First Star in the Universe}.
\newblock \emph{Science}, 295:\penalty0 93--98, January 2002.
\newblock \doi{10.1126/science.295.5552.93}.

\bibitem[{Abitbol} et~al.(2017){Abitbol}, {Ahmed}, {Barron}, {Basu Thakur},
  {Bender}, {Benson}, {Bischoff}, {Bryan}, {Carlstrom}, {Chang}, {Chuss},
  {Cukierman}, {de Haan}, {Dobbs}, {Essinger-Hileman}, {Filippini}, {Ganga},
  {Gudmundsson}, {Halverson}, {Hanany}, {Henderson}, {Hill}, {Ho}, {Hubmayr},
  {Irwin}, {Jeong}, {Johnson}, {Kernasovskiy}, {Kovac}, {Kusaka}, {Lee},
  {Maria}, {Mauskopf}, {McMahon}, {Moncelsi}, {Nadolski}, {Nagy}, {Niemack},
  {O'Brient}, {Padin}, {Parshley}, {Pryke}, {Roe}, {Rostem}, {Ruhl}, {Simon},
  {Staggs}, {Suzuki}, {Switzer}, {Thompson}, {Timbie}, {Tucker}, {Vieira},
  {Vieregg}, {Westbrook}, {Wollack}, {Yoon}, {Young}, and
  {Young}]{Abitbol:2017}
M.~H. {Abitbol}, Z.~{Ahmed}, D.~{Barron}, R.~{Basu Thakur}, A.~N. {Bender},
  B.~A. {Benson}, C.~A. {Bischoff}, S.~A. {Bryan}, J.~E. {Carlstrom}, C.~L.
  {Chang}, D.~T. {Chuss}, A.~{Cukierman}, T.~{de Haan}, M.~{Dobbs},
  T.~{Essinger-Hileman}, J.~P. {Filippini}, K.~{Ganga}, J.~E. {Gudmundsson},
  N.~W. {Halverson}, S.~{Hanany}, S.~W. {Henderson}, C.~A. {Hill}, S.-P.~P.
  {Ho}, J.~{Hubmayr}, K.~{Irwin}, O.~{Jeong}, B.~R. {Johnson}, S.~A.
  {Kernasovskiy}, J.~M. {Kovac}, A.~{Kusaka}, A.~T. {Lee}, S.~{Maria},
  P.~{Mauskopf}, J.~J. {McMahon}, L.~{Moncelsi}, A.~W. {Nadolski}, J.~M.
  {Nagy}, M.~D. {Niemack}, R.~C. {O'Brient}, S.~{Padin}, S.~C. {Parshley},
  C.~{Pryke}, N.~A. {Roe}, K.~{Rostem}, J.~{Ruhl}, S.~M. {Simon}, S.~T.
  {Staggs}, A.~{Suzuki}, E.~R. {Switzer}, K.~L. {Thompson}, P.~{Timbie}, G.~S.
  {Tucker}, J.~D. {Vieira}, A.~G. {Vieregg}, B.~{Westbrook}, E.~J. {Wollack},
  K.~W. {Yoon}, K.~S. {Young}, and E.~Y. {Young}.
\newblock {CMB-S4 Technology Book, First Edition}.
\newblock \emph{ArXiv e-prints}, June 2017.

\bibitem[{Agarwal} et~al.(2012){Agarwal}, {Khochfar}, {Johnson}, {Neistein},
  {Dalla Vecchia}, and {Livio}]{Agarwal:2012}
B.~{Agarwal}, S.~{Khochfar}, J.~L. {Johnson}, E.~{Neistein}, C.~{Dalla
  Vecchia}, and M.~{Livio}.
\newblock {Ubiquitous seeding of supermassive black holes by direct collapse}.
\newblock \emph{\mnras}, 425:\penalty0 2854--2871, October 2012.
\newblock \doi{10.1111/j.1365-2966.2012.21651.x}.

\bibitem[{Arendt} et~al.(2000){Arendt}, {Fixsen}, and {Moseley}]{Arendt:2000}
R.~G. {Arendt}, D.~J. {Fixsen}, and S.~H. {Moseley}.
\newblock {Dithering Strategies for Efficient Self-Calibration of Imaging
  Arrays}.
\newblock \emph{\apj}, 536:\penalty0 500--512, June 2000.
\newblock \doi{10.1086/308923}.

\bibitem[{Arendt} et~al.(2002){Arendt}, {Fixsen}, and {Moseley}]{Arendt:2002}
R.~G. {Arendt}, D.~J. {Fixsen}, and S.~H. {Moseley}.
\newblock {A Practical Demonstration of Self-Calibration of NICMOS HDF North
  and South Data}.
\newblock In D.~A. {Bohlender}, D.~{Durand}, and T.~H. {Handley}, editors,
  \emph{Astronomical Data Analysis Software and Systems XI}, volume 281 of
  \emph{Astronomical Society of the Pacific Conference Series}, page 217, 2002.

\bibitem[{Arendt} et~al.(2010){Arendt}, {Kashlinsky}, {Moseley}, and
  {Mather}]{Arendt:2010}
R.~G. {Arendt}, A.~{Kashlinsky}, S.~H. {Moseley}, and J.~{Mather}.
\newblock {Cosmic Infrared Background Fluctuations in Deep Spitzer Infrared
  Array Camera Images: Data Processing and Analysis}.
\newblock \emph{\apjs}, 186:\penalty0 10--47, January 2010.
\newblock \doi{10.1088/0067-0049/186/1/10}.

\bibitem[{Arendt} et~al.(2016){Arendt}, {Kashlinsky}, {Moseley}, and
  {Mather}]{Arendt:2016}
R.~G. {Arendt}, A.~{Kashlinsky}, S.~H. {Moseley}, and J.~{Mather}.
\newblock {Cosmic Infrared Background Fluctuations and Zodiacal Light}.
\newblock \emph{\apj}, 824:\penalty0 26, June 2016.
\newblock \doi{10.3847/0004-637X/824/1/26}.

\bibitem[{Atrio-Barandela} and {Kashlinsky}(2014)]{Atrio-Barandela:2014}
F.~{Atrio-Barandela} and A.~{Kashlinsky}.
\newblock {Probing the Epoch of Pre-reionization by Cross-correlating Cosmic
  Microwave and Infrared Background Anisotropies}.
\newblock \emph{\apjl}, 797:\penalty0 L26, December 2014.
\newblock \doi{10.1088/2041-8205/797/2/L26}.

\bibitem[{Bond} and {Efstathiou}(1984)]{Bond:1984}
J.~R. {Bond} and G.~{Efstathiou}.
\newblock {Cosmic background radiation anisotropies in universes dominated by
  nonbaryonic dark matter}.
\newblock \emph{\apjl}, 285:\penalty0 L45--L48, October 1984.
\newblock \doi{10.1086/184362}.

\bibitem[{Bromm} and {Larson}(2004)]{Bromm:2004}
V.~{Bromm} and R.~B. {Larson}.
\newblock {The First Stars}.
\newblock \emph{\araa}, 42:\penalty0 79--118, September 2004.
\newblock \doi{10.1146/annurev.astro.42.053102.134034}.

\bibitem[{Bromm} et~al.(2001){Bromm}, {Ferrara}, {Coppi}, and
  {Larson}]{Bromm:2001}
V.~{Bromm}, A.~{Ferrara}, P.~S. {Coppi}, and R.~B. {Larson}.
\newblock {The fragmentation of pre-enriched primordial objects}.
\newblock \emph{\mnras}, 328:\penalty0 969--976, December 2001.
\newblock \doi{10.1046/j.1365-8711.2001.04915.x}.

\bibitem[{Cappelluti} et~al.(2013){Cappelluti}, {Kashlinsky}, {Arendt},
  {Comastri}, {Fazio}, {Finoguenov}, {Hasinger}, {Mather}, {Miyaji}, and
  {Moseley}]{Cappelluti:2013}
N.~{Cappelluti}, A.~{Kashlinsky}, R.~G. {Arendt}, A.~{Comastri}, G.~G. {Fazio},
  A.~{Finoguenov}, G.~{Hasinger}, J.~C. {Mather}, T.~{Miyaji}, and S.~H.
  {Moseley}.
\newblock {Cross-correlating Cosmic Infrared and X-Ray Background Fluctuations:
  Evidence of Significant Black Hole Populations among the CIB Sources}.
\newblock \emph{\apj}, 769:\penalty0 68, May 2013.
\newblock \doi{10.1088/0004-637X/769/1/68}.

\bibitem[{Cappelluti} et~al.(2017){Cappelluti}, {Arendt}, {Kashlinsky}, {Li},
  {Hasinger}, {Helgason}, {Urry}, {Natarajan}, and
  {Finoguenov}]{Cappelluti:2017}
N.~{Cappelluti}, R.~{Arendt}, A.~{Kashlinsky}, Y.~{Li}, G.~{Hasinger},
  K.~{Helgason}, M.~{Urry}, P.~{Natarajan}, and A.~{Finoguenov}.
\newblock {Probing Large-scale Coherence between Spitzer IR and Chandra X-Ray
  Source-subtracted Cosmic Backgrounds}.
\newblock \emph{\apjl}, 847:\penalty0 L11, September 2017.
\newblock \doi{10.3847/2041-8213/aa8acd}.

\bibitem[{Cooray} et~al.(2004){Cooray}, {Bock}, {Keatin}, {Lange}, and
  {Matsumoto}]{Cooray:2004}
A.~{Cooray}, J.~J. {Bock}, B.~{Keatin}, A.~E. {Lange}, and T.~{Matsumoto}.
\newblock {First Star Signature in Infrared Background Anisotropies}.
\newblock \emph{\apj}, 606:\penalty0 611--624, May 2004.
\newblock \doi{10.1086/383137}.

\bibitem[{Cooray} et~al.(2012{\natexlab{a}}){Cooray}, {Gong}, {Smidt}, and
  {Santos}]{Cooray:2012a}
A.~{Cooray}, Y.~{Gong}, J.~{Smidt}, and M.~G. {Santos}.
\newblock {The Near-infrared Background Intensity and Anisotropies during the
  Epoch of Reionization}.
\newblock \emph{\apj}, 756:\penalty0 92, September 2012{\natexlab{a}}.
\newblock \doi{10.1088/0004-637X/756/1/92}.

\bibitem[{Cooray} et~al.(2012{\natexlab{b}}){Cooray}, {Smidt}, {de Bernardis},
  {Gong}, {Stern}, {Ashby}, {Eisenhardt}, {Frazer}, {Gonzalez}, {Kochanek},
  {Koz{\l}owski}, and {Wright}]{Cooray:2012}
A.~{Cooray}, J.~{Smidt}, F.~{de Bernardis}, Y.~{Gong}, D.~{Stern}, M.~L.~N.
  {Ashby}, P.~R. {Eisenhardt}, C.~C. {Frazer}, A.~H. {Gonzalez}, C.~S.
  {Kochanek}, S.~{Koz{\l}owski}, and E.~L. {Wright}.
\newblock {Near-infrared background anisotropies from diffuse intrahalo light
  of galaxies}.
\newblock \emph{\nat}, 490:\penalty0 514--516, October 2012{\natexlab{b}}.
\newblock \doi{10.1038/nature11474}.

\bibitem[{Driver} et~al.(2016){Driver}, {Andrews}, {Davies}, {Robotham},
  {Wright}, {Windhorst}, {Cohen}, {Emig}, {Jansen}, and {Dunne}]{Driver:2016}
S.~P. {Driver}, S.~K. {Andrews}, L.~J. {Davies}, A.~S.~G. {Robotham}, A.~H.
  {Wright}, R.~A. {Windhorst}, S.~{Cohen}, K.~{Emig}, R.~A. {Jansen}, and
  L.~{Dunne}.
\newblock {Measurements of Extragalactic Background Light from the Far UV to
  the Far IR from Deep Ground- and Space-based Galaxy Counts}.
\newblock \emph{\apj}, 827:\penalty0 108, August 2016.
\newblock \doi{10.3847/0004-637X/827/2/108}.

\bibitem[{Fermi-LAT Collaboration} et~al.(2018){Fermi-LAT Collaboration},
  {Abdollahi}, {Ackermann}, {Ajello}, {Atwood}, {Baldini}, {Ballet},
  {Barbiellini}, {Bastieri}, {Becerra Gonzalez}, {Bellazzini}, {Bissaldi},
  {Blandford}, {Bloom}, {Bonino}, {Bottacini}, {Buson}, {Bregeon}, {Bruel},
  {Buehler}, {Cameron}, {Caputo}, {Caraveo}, {Cavazzuti}, {Charles}, {Chen},
  {Cheung}, {Chiaro}, {Ciprini}, {Cohen-Tanugi}, {Cominsky}, {Conrad},
  {Costantin}, {Cutini}, {D'Ammando}, {de Palma}, {Desai}, {Digel}, {Di Lalla},
  {Di Mauro}, {Di Venere}, {Dom{\'{\i}}nguez}, {Favuzzi}, {Fegan}, {Finke},
  {Franckowiak}, {Fukazawa}, {Funk}, {Fusco}, {Gallardo Romero}, {Gargano},
  {Gasparrini}, {Giglietto}, {Giordano}, {Giroletti}, {Green}, {Grenier},
  {Guillemot}, {Guiriec}, {Hartmann}, {Hays}, {Helgason}, {Horan},
  {J{\'o}hannesson}, {Kocevski}, {Kuss}, {Larsson}, {Latronico}, {Li}, {Longo},
  {Loparco}, {Lott}, {Lovellette}, {Lubrano}, {Madejski}, {Magill}, {Maldera},
  {Manfreda}, {Marcotulli}, {Mazziotta}, {McEnery}, {Meyer}, {Michelson},
  {Mizuno}, {Monzani}, {Morselli}, {Moskalenko}, {Negro}, {Nuss}, {Ojha},
  {Omodei}, {Orienti}, {Orlando}, {Ormes}, {Palatiello}, {Paliya}, {Paneque},
  {Perkins}, {Persic}, {Pesce-Rollins}, {Petrosian}, {Piron}, {Porter},
  {Primack}, {Principe}, {Rain{\`o}}, {Rando}, {Razzano}, {Razzaque}, {Reimer},
  {Reimer}, {Saz Parkinson}, {Sgr{\`o}}, {Siskind}, {Spandre}, {Spinelli},
  {Suson}, {Tajima}, {Takahashi}, {Thayer}, {Tibaldo}, {Torres}, {Torresi},
  {Tosti}, {Tramacere}, {Troja}, {Valverde}, {Vianello}, {Vogel}, {Wood}, and
  {Zaharijas}]{Fermi-LAT-Collaboration:2018}
{Fermi-LAT Collaboration}, S.~{Abdollahi}, M.~{Ackermann}, M.~{Ajello}, W.~B.
  {Atwood}, L.~{Baldini}, J.~{Ballet}, G.~{Barbiellini}, D.~{Bastieri},
  J.~{Becerra Gonzalez}, R.~{Bellazzini}, E.~{Bissaldi}, R.~D. {Blandford},
  E.~D. {Bloom}, R.~{Bonino}, E.~{Bottacini}, S.~{Buson}, J.~{Bregeon},
  P.~{Bruel}, R.~{Buehler}, R.~A. {Cameron}, R.~{Caputo}, P.~A. {Caraveo},
  E.~{Cavazzuti}, E.~{Charles}, S.~{Chen}, C.~C. {Cheung}, G.~{Chiaro},
  S.~{Ciprini}, J.~{Cohen-Tanugi}, L.~R. {Cominsky}, J.~{Conrad},
  D.~{Costantin}, S.~{Cutini}, F.~{D'Ammando}, F.~{de Palma}, A.~{Desai}, S.~W.
  {Digel}, N.~{Di Lalla}, M.~{Di Mauro}, L.~{Di Venere}, A.~{Dom{\'{\i}}nguez},
  C.~{Favuzzi}, S.~J. {Fegan}, J.~{Finke}, A.~{Franckowiak}, Y.~{Fukazawa},
  S.~{Funk}, P.~{Fusco}, G.~{Gallardo Romero}, F.~{Gargano}, D.~{Gasparrini},
  N.~{Giglietto}, F.~{Giordano}, M.~{Giroletti}, D.~{Green}, I.~A. {Grenier},
  L.~{Guillemot}, S.~{Guiriec}, D.~H. {Hartmann}, E.~{Hays}, K.~{Helgason},
  D.~{Horan}, G.~{J{\'o}hannesson}, D.~{Kocevski}, M.~{Kuss}, S.~{Larsson},
  L.~{Latronico}, J.~{Li}, F.~{Longo}, F.~{Loparco}, B.~{Lott}, M.~N.
  {Lovellette}, P.~{Lubrano}, G.~M. {Madejski}, J.~D. {Magill}, S.~{Maldera},
  A.~{Manfreda}, L.~{Marcotulli}, M.~N. {Mazziotta}, J.~E. {McEnery},
  M.~{Meyer}, P.~F. {Michelson}, T.~{Mizuno}, M.~E. {Monzani}, A.~{Morselli},
  I.~V. {Moskalenko}, M.~{Negro}, E.~{Nuss}, R.~{Ojha}, N.~{Omodei},
  M.~{Orienti}, E.~{Orlando}, J.~F. {Ormes}, M.~{Palatiello}, V.~S. {Paliya},
  D.~{Paneque}, J.~S. {Perkins}, M.~{Persic}, M.~{Pesce-Rollins},
  V.~{Petrosian}, F.~{Piron}, T.~A. {Porter}, J.~R. {Primack}, G.~{Principe},
  S.~{Rain{\`o}}, R.~{Rando}, M.~{Razzano}, S.~{Razzaque}, A.~{Reimer},
  O.~{Reimer}, P.~M. {Saz Parkinson}, C.~{Sgr{\`o}}, E.~J. {Siskind},
  G.~{Spandre}, P.~{Spinelli}, D.~J. {Suson}, H.~{Tajima}, M.~{Takahashi},
  J.~B. {Thayer}, L.~{Tibaldo}, D.~F. {Torres}, E.~{Torresi}, G.~{Tosti},
  A.~{Tramacere}, E.~{Troja}, J.~{Valverde}, G.~{Vianello}, M.~{Vogel},
  K.~{Wood}, and G.~{Zaharijas}.
\newblock {A gamma-ray determination of the Universe's star formation history}.
\newblock \emph{Science}, 362:\penalty0 1031--1034, November 2018.
\newblock \doi{10.1126/science.aat8123}.

\bibitem[{Fixsen} et~al.(2000){Fixsen}, {Moseley}, and {Arendt}]{Fixsen:2000}
D.~J. {Fixsen}, S.~H. {Moseley}, and R.~G. {Arendt}.
\newblock {Calibrating Array Detectors}.
\newblock \emph{\apjs}, 128:\penalty0 651--658, June 2000.
\newblock \doi{10.1086/313390}.

\bibitem[{Gong} et~al.(2015){Gong}, {Cooray}, {Mitchell-Wynne}, {Chen},
  {Zemcov}, and {Smidt}]{Gong:2015}
Y.~{Gong}, A.~{Cooray}, K.~{Mitchell-Wynne}, X.~{Chen}, M.~{Zemcov}, and
  J.~{Smidt}.
\newblock {Axion decay and anisotropy of near-IR extragalactic background
  light}.
\newblock \emph{ArXiv e-prints}, November 2015.

\bibitem[{Hauser} et~al.(1998){Hauser}, {Arendt}, {Kelsall}, {Dwek}, {Odegard},
  {Weiland}, {Freudenreich}, {Reach}, {Silverberg}, {Moseley}, {Pei}, {Lubin},
  {Mather}, {Shafer}, {Smoot}, {Weiss}, {Wilkinson}, and {Wright}]{Hauser:1998}
M.~G. {Hauser}, R.~G. {Arendt}, T.~{Kelsall}, E.~{Dwek}, N.~{Odegard}, J.~L.
  {Weiland}, H.~T. {Freudenreich}, W.~T. {Reach}, R.~F. {Silverberg}, S.~H.
  {Moseley}, Y.~C. {Pei}, P.~{Lubin}, J.~C. {Mather}, R.~A. {Shafer}, G.~F.
  {Smoot}, R.~{Weiss}, D.~T. {Wilkinson}, and E.~L. {Wright}.
\newblock {The COBE Diffuse Infrared Background Experiment Search for the
  Cosmic Infrared Background. I. Limits and Detections}.
\newblock \emph{\apj}, 508:\penalty0 25--43, November 1998.
\newblock \doi{10.1086/306379}.

\bibitem[{Helgason} et~al.(2012){Helgason}, {Ricotti}, and
  {Kashlinsky}]{Helgason:2012a}
K.~{Helgason}, M.~{Ricotti}, and A.~{Kashlinsky}.
\newblock {Reconstructing the Near-infrared Background Fluctuations from Known
  Galaxy Populations Using Multiband Measurements of Luminosity Functions}.
\newblock \emph{\apj}, 752:\penalty0 113, June 2012.
\newblock \doi{10.1088/0004-637X/752/2/113}.

\bibitem[{Helgason} et~al.(2014){Helgason}, {Cappelluti}, {Hasinger},
  {Kashlinsky}, and {Ricotti}]{Helgason:2014}
K.~{Helgason}, N.~{Cappelluti}, G.~{Hasinger}, A.~{Kashlinsky}, and
  M.~{Ricotti}.
\newblock {The Contribution of $z \lesssim 6$ Sources to the Spatial Coherence
  in the Unresolved Cosmic Near-infrared and X-Ray Backgrounds}.
\newblock \emph{\apj}, 785:\penalty0 38, April 2014.
\newblock \doi{10.1088/0004-637X/785/1/38}.

\bibitem[{Helgason} et~al.(2016){Helgason}, {Ricotti}, {Kashlinsky}, and
  {Bromm}]{Helgason:2016}
K.~{Helgason}, M.~{Ricotti}, A.~{Kashlinsky}, and V.~{Bromm}.
\newblock {On the physical requirements for a pre-reionization origin of the
  unresolved near-infrared background}.
\newblock \emph{\mnras}, 455:\penalty0 282--294, January 2016.
\newblock \doi{10.1093/mnras/stv2209}.

\bibitem[{Kashlinsky}(2005)]{Kashlinsky:2005}
A.~{Kashlinsky}.
\newblock {Cosmic infrared background and early galaxy evolution [review
  article]}.
\newblock \emph{\physrep}, 409:\penalty0 361--438, April 2005.
\newblock \doi{10.1016/j.physrep.2004.12.005}.

\bibitem[{Kashlinsky}(2016)]{Kashlinsky:2016}
A.~{Kashlinsky}.
\newblock {LIGO Gravitational Wave Detection, Primordial Black Holes, and the
  Near-IR Cosmic Infrared Background Anisotropies}.
\newblock \emph{\apjl}, 823:\penalty0 L25, June 2016.
\newblock \doi{10.3847/2041-8205/823/2/L25}.

\bibitem[{Kashlinsky} and {Odenwald}(2000)]{Kashlinsky:2000}
A.~{Kashlinsky} and S.~{Odenwald}.
\newblock {Clustering of the Diffuse Infrared Light from the COBE DIRBE Maps.
  III. Power Spectrum Analysis and Excess Isotropic Component of Fluctuations}.
\newblock \emph{\apj}, 528:\penalty0 74--95, January 2000.
\newblock \doi{10.1086/308172}.

\bibitem[{Kashlinsky} et~al.(1996{\natexlab{a}}){Kashlinsky}, {Mather}, and
  {Odenwald}]{Kashlinsky:1996}
A.~{Kashlinsky}, J.~C. {Mather}, and S.~{Odenwald}.
\newblock {Clustering of the Diffuse Infrared Light from the COBE DIRBE Maps:
  an All-Sky Survey of C(0)}.
\newblock \emph{\apjl}, 473:\penalty0 L9, December 1996{\natexlab{a}}.
\newblock \doi{10.1086/310379}.

\bibitem[{Kashlinsky} et~al.(1996{\natexlab{b}}){Kashlinsky}, {Mather},
  {Odenwald}, and {Hauser}]{Kashlinsky:1996a}
A.~{Kashlinsky}, J.~C. {Mather}, S.~{Odenwald}, and M.~G. {Hauser}.
\newblock {Clustering of the Diffuse Infrared Light from the COBE DIRBE Maps.
  I. C(0) and Limits on the Near-Infrared Background}.
\newblock \emph{\apj}, 470:\penalty0 681, October 1996{\natexlab{b}}.
\newblock \doi{10.1086/177900}.

\bibitem[{Kashlinsky} et~al.(2002){Kashlinsky}, {Odenwald}, {Mather},
  {Skrutskie}, and {Cutri}]{Kashlinsky:2002}
A.~{Kashlinsky}, S.~{Odenwald}, J.~{Mather}, M.~F. {Skrutskie}, and R.~M.
  {Cutri}.
\newblock {Detection of Small-Scale Fluctuations in the Near-Infrared Cosmic
  Infrared Background from Long-Exposure 2MASS Fields}.
\newblock \emph{\apjl}, 579:\penalty0 L53--L57, November 2002.
\newblock \doi{10.1086/345335}.

\bibitem[{Kashlinsky} et~al.(2004){Kashlinsky}, {Arendt}, {Gardner}, {Mather},
  and {Moseley}]{Kashlinsky:2004}
A.~{Kashlinsky}, R.~{Arendt}, J.~P. {Gardner}, J.~C. {Mather}, and S.~H.
  {Moseley}.
\newblock {Detecting Population III Stars through Observations of Near-Infrared
  Cosmic Infrared Background Anisotropies}.
\newblock \emph{\apj}, 608:\penalty0 1--9, June 2004.
\newblock \doi{10.1086/386365}.

\bibitem[{Kashlinsky} et~al.(2005){Kashlinsky}, {Arendt}, {Mather}, and
  {Moseley}]{Kashlinsky:2005a}
A.~{Kashlinsky}, R.~G. {Arendt}, J.~{Mather}, and S.~H. {Moseley}.
\newblock {Tracing the first stars with fluctuations of the cosmic infrared
  background}.
\newblock \emph{\nat}, 438:\penalty0 45--50, November 2005.
\newblock \doi{10.1038/nature04143}.

\bibitem[{Kashlinsky} et~al.(2007{\natexlab{a}}){Kashlinsky}, {Arendt},
  {Mather}, and {Moseley}]{Kashlinsky:2007}
A.~{Kashlinsky}, R.~G. {Arendt}, J.~{Mather}, and S.~H. {Moseley}.
\newblock {Demonstrating the Negligible Contribution of Optical HST ACS
  Galaxies to Source-subtracted Cosmic Infrared Background Fluctuations in Deep
  Spitzer IRAC Images}.
\newblock \emph{\apjl}, 666:\penalty0 L1--L4, September 2007{\natexlab{a}}.
\newblock \doi{10.1086/521551}.

\bibitem[{Kashlinsky} et~al.(2007{\natexlab{b}}){Kashlinsky}, {Arendt},
  {Mather}, and {Moseley}]{Kashlinsky:2007a}
A.~{Kashlinsky}, R.~G. {Arendt}, J.~{Mather}, and S.~H. {Moseley}.
\newblock {New Measurements of Cosmic Infrared Background Fluctuations from
  Early Epochs}.
\newblock \emph{\apjl}, 654:\penalty0 L5--L8, January 2007{\natexlab{b}}.
\newblock \doi{10.1086/510483}.

\bibitem[{Kashlinsky} et~al.(2007{\natexlab{c}}){Kashlinsky}, {Arendt},
  {Mather}, and {Moseley}]{Kashlinsky:2007b}
A.~{Kashlinsky}, R.~G. {Arendt}, J.~{Mather}, and S.~H. {Moseley}.
\newblock {On the Nature of the Sources of the Cosmic Infrared Background}.
\newblock \emph{\apjl}, 654:\penalty0 L1--L4, January 2007{\natexlab{c}}.
\newblock \doi{10.1086/510484}.

\bibitem[{Kashlinsky} et~al.(2012){Kashlinsky}, {Arendt}, {Ashby}, {Fazio},
  {Mather}, and {Moseley}]{Kashlinsky:2012}
A.~{Kashlinsky}, R.~G. {Arendt}, M.~L.~N. {Ashby}, G.~G. {Fazio}, J.~{Mather},
  and S.~H. {Moseley}.
\newblock {New Measurements of the Cosmic Infrared Background Fluctuations in
  Deep Spitzer/IRAC Survey Data and Their Cosmological Implications}.
\newblock \emph{\apj}, 753:\penalty0 63, July 2012.
\newblock \doi{10.1088/0004-637X/753/1/63}.

\bibitem[{Kashlinsky} et~al.(2015{\natexlab{a}}){Kashlinsky}, {Arendt},
  {Atrio-Barandela}, and {Helgason}]{Kashlinsky:2015}
A.~{Kashlinsky}, R.~G. {Arendt}, F.~{Atrio-Barandela}, and K.~{Helgason}.
\newblock {Lyman-tomography of Cosmic Infrared Background Fluctuations with
  Euclid: Probing Emissions and Baryonic Acoustic Oscillations at $z \gtrsim
  10$}.
\newblock \emph{\apjl}, 813:\penalty0 L12, November 2015{\natexlab{a}}.
\newblock \doi{10.1088/2041-8205/813/1/L12}.

\bibitem[{Kashlinsky} et~al.(2015{\natexlab{b}}){Kashlinsky}, {Mather},
  {Helgason}, {Arendt}, {Bromm}, and {Moseley}]{Kashlinsky:2015a}
A.~{Kashlinsky}, J.~C. {Mather}, K.~{Helgason}, R.~G. {Arendt}, V.~{Bromm}, and
  S.~H. {Moseley}.
\newblock {Reconstructing Emission from Pre-reionization Sources with Cosmic
  Infrared Background Fluctuation Measurements by the JWST}.
\newblock \emph{\apj}, 804:\penalty0 99, May 2015{\natexlab{b}}.
\newblock \doi{10.1088/0004-637X/804/2/99}.

\bibitem[{Kashlinsky} et~al.(2018){Kashlinsky}, {Arendt}, {Atrio-Barandela},
  {Cappelluti}, {Ferrara}, and {Hasinger}]{Kashlinsky:2018}
A.~{Kashlinsky}, R.~G. {Arendt}, F.~{Atrio-Barandela}, N.~{Cappelluti},
  A.~{Ferrara}, and G.~{Hasinger}.
\newblock {Looking at cosmic near-infrared background radiation anisotropies}.
\newblock \emph{Reviews of Modern Physics}, 90\penalty0 (2):\penalty0 025006,
  April 2018.
\newblock \doi{10.1103/RevModPhys.90.025006}.

\bibitem[{Kashlinsky} et~al.(2019){Kashlinsky}, {Arendt}, {Cappelluti},
  {Finoguenov}, {Hasinger}, {Helgason}, and {Merloni}]{Kashlinsky:2019}
A.~{Kashlinsky}, R.~G. {Arendt}, N.~{Cappelluti}, A.~{Finoguenov},
  G.~{Hasinger}, K.~{Helgason}, and A.~{Merloni}.
\newblock {Probing the Cross-power of Unresolved Cosmic Infrared and X-Ray
  Backgrounds with Upcoming Space Missions}.
\newblock \emph{\apjl}, 871:\penalty0 L6, January 2019.
\newblock \doi{10.3847/2041-8213/aafaf6}.

\bibitem[{Latif} and {Ferrara}(2016)]{Latif:2016}
M.~A. {Latif} and A.~{Ferrara}.
\newblock {Formation of Supermassive Black Hole Seeds}.
\newblock \emph{\pasa}, 33:\penalty0 e051, October 2016.
\newblock \doi{10.1017/pasa.2016.41}.

\bibitem[{Laureijs} et~al.(2011){Laureijs}, {Amiaux}, {Arduini},
  {Augu{\`e}res}, {Brinchmann}, {Cole}, {Cropper}, {Dabin}, {Duvet}, {Ealet},
  and et~al.]{Laureijs:2011}
R.~{Laureijs}, J.~{Amiaux}, S.~{Arduini}, J.~. {Augu{\`e}res}, J.~{Brinchmann},
  R.~{Cole}, M.~{Cropper}, C.~{Dabin}, L.~{Duvet}, A.~{Ealet}, and et~al.
\newblock {Euclid Definition Study Report}.
\newblock \emph{ArXiv e-prints}, October 2011.

\bibitem[{Laureijs} et~al.(2014){Laureijs}, {Racca}, {Stagnaro}, {Salvignol},
  {Lorenzo Alvarez}, {Saavedra Criado}, {Gaspar Venancio}, {Short}, {Strada},
  {Colombo}, {Buenadicha}, {Hoar}, {Kohley}, {Vavrek}, {Mellier}, {Berthe},
  {Amiaux}, {Cropper}, {Niemi}, {Pottinger}, {Ealet}, {Jahnke}, {Maciaszek},
  {Pasian}, {Sauvage}, {Wachter}, {Israelsson}, {Holmes}, {Seiffert},
  {Cazaubiel}, {Anselmi}, and {Musi}]{Laureijs:2014}
R.~{Laureijs}, G.~{Racca}, L.~{Stagnaro}, J.-C. {Salvignol}, J.~{Lorenzo
  Alvarez}, G.~{Saavedra Criado}, L.~{Gaspar Venancio}, A.~{Short},
  P.~{Strada}, C.~{Colombo}, G.~{Buenadicha}, J.~{Hoar}, R.~{Kohley},
  R.~{Vavrek}, Y.~{Mellier}, M.~{Berthe}, J.~{Amiaux}, M.~{Cropper},
  S.~{Niemi}, S.~{Pottinger}, A.~{Ealet}, K.~{Jahnke}, T.~{Maciaszek},
  F.~{Pasian}, M.~{Sauvage}, S.~{Wachter}, U.~{Israelsson}, W.~{Holmes},
  M.~{Seiffert}, V.~{Cazaubiel}, A.~{Anselmi}, and P.~{Musi}.
\newblock {Euclid mission status}.
\newblock In \emph{Space Telescopes and Instrumentation 2014: Optical,
  Infrared, and Millimeter Wave}, volume 9143 of \emph{\procspie}, page 91430H,
  August 2014.
\newblock \doi{10.1117/12.2054883}.

\bibitem[{Li} et~al.(2018){Li}, {Cappelluti}, {Arendt}, {Hasinger},
  {Kashlinsky}, and {Helgason}]{Li:2018}
Y.~{Li}, N.~{Cappelluti}, R.~G. {Arendt}, G.~{Hasinger}, A.~{Kashlinsky}, and
  K.~{Helgason}.
\newblock {The SPLASH and Chandra COSMOS Legacy Survey: The Cross-power between
  Near-infrared and X-Ray Background Fluctuations}.
\newblock \emph{\apj}, 864:\penalty0 141, September 2018.
\newblock \doi{10.3847/1538-4357/aad55a}.

\bibitem[{Matsumoto} et~al.(2005){Matsumoto}, {Matsuura}, {Murakami}, {Tanaka},
  {Freund}, {Lim}, {Cohen}, {Kawada}, and {Noda}]{Matsumoto:2005}
T.~{Matsumoto}, S.~{Matsuura}, H.~{Murakami}, M.~{Tanaka}, M.~{Freund},
  M.~{Lim}, M.~{Cohen}, M.~{Kawada}, and M.~{Noda}.
\newblock {Infrared Telescope in Space Observations of the Near-Infrared
  Extragalactic Background Light}.
\newblock \emph{\apj}, 626:\penalty0 31--43, June 2005.
\newblock \doi{10.1086/429383}.

\bibitem[{Matsumoto} et~al.(2011){Matsumoto}, {Seo}, {Jeong}, {Lee},
  {Matsuura}, {Matsuhara}, {Oyabu}, {Pyo}, and {Wada}]{Matsumoto:2011}
T.~{Matsumoto}, H.~J. {Seo}, W.-S. {Jeong}, H.~M. {Lee}, S.~{Matsuura},
  H.~{Matsuhara}, S.~{Oyabu}, J.~{Pyo}, and T.~{Wada}.
\newblock {AKARI Observation of the Fluctuation of the Near-infrared
  Background}.
\newblock \emph{\apj}, 742:\penalty0 124, December 2011.
\newblock \doi{10.1088/0004-637X/742/2/124}.

\bibitem[{Merloni} et~al.(2012){Merloni}, {Predehl}, {Becker}, {B{\"o}hringer},
  {Boller}, {Brunner}, {Brusa}, {Dennerl}, {Freyberg}, {Friedrich},
  {Georgakakis}, {Haberl}, {Hasinger}, {Meidinger}, {Mohr}, {Nandra}, {Rau},
  {Reiprich}, {Robrade}, {Salvato}, {Santangelo}, {Sasaki}, {Schwope}, {Wilms},
  and {German eROSITA Consortium}]{Merloni:2012}
A.~{Merloni}, P.~{Predehl}, W.~{Becker}, H.~{B{\"o}hringer}, T.~{Boller},
  H.~{Brunner}, M.~{Brusa}, K.~{Dennerl}, M.~{Freyberg}, P.~{Friedrich},
  A.~{Georgakakis}, F.~{Haberl}, G.~{Hasinger}, N.~{Meidinger}, J.~{Mohr},
  K.~{Nandra}, A.~{Rau}, T.~H. {Reiprich}, J.~{Robrade}, M.~{Salvato},
  A.~{Santangelo}, M.~{Sasaki}, A.~{Schwope}, J.~{Wilms}, and t.~{German
  eROSITA Consortium}.
\newblock {eROSITA Science Book: Mapping the Structure of the Energetic
  Universe}.
\newblock \emph{ArXiv e-prints}, September 2012.

\bibitem[{Mesinger}(2016)]{Mesinger:2016}
A.~{Mesinger}, editor.
\newblock \emph{{Understanding the Epoch of Cosmic Reionization}}, volume 423
  of \emph{Astrophysics and Space Science Library}, 2016.
\newblock \doi{10.1007/978-3-319-21957-8}.

\bibitem[{Mitchell-Wynne} et~al.(2015){Mitchell-Wynne}, {Cooray}, {Gong},
  {Ashby}, {Dolch}, {Ferguson}, {Finkelstein}, {Grogin}, {Kocevski},
  {Koekemoer}, {Primack}, and {Smidt}]{Mitchell-Wynne:2015}
K.~{Mitchell-Wynne}, A.~{Cooray}, Y.~{Gong}, M.~{Ashby}, T.~{Dolch},
  H.~{Ferguson}, S.~{Finkelstein}, N.~{Grogin}, D.~{Kocevski}, A.~{Koekemoer},
  J.~{Primack}, and J.~{Smidt}.
\newblock {Ultraviolet luminosity density of the universe during the epoch of
  reionization}.
\newblock \emph{Nature Communications}, 6:\penalty0 7945, September 2015.
\newblock \doi{10.1038/ncomms8945}.

\bibitem[{Mitchell-Wynne} et~al.(2016){Mitchell-Wynne}, {Cooray}, {Xue}, {Luo},
  {Brandt}, and {Koekemoer}]{Mitchell-Wynne:2016}
K.~{Mitchell-Wynne}, A.~{Cooray}, Y.~{Xue}, B.~{Luo}, W.~{Brandt}, and
  A.~{Koekemoer}.
\newblock {Cross-correlation between X-ray and optical/near-infrared background
  intensity fluctuations}.
\newblock \emph{ArXiv e-prints}, October 2016.

\bibitem[{Nandra} et~al.(2013){Nandra}, {Barret}, {Barcons}, {Fabian}, {den
  Herder}, {Piro}, {Watson}, {Adami}, {Aird}, {Afonso}, and
  et~al.]{Nandra:2013}
K.~{Nandra}, D.~{Barret}, X.~{Barcons}, A.~{Fabian}, J.-W. {den Herder},
  L.~{Piro}, M.~{Watson}, C.~{Adami}, J.~{Aird}, J.~M. {Afonso}, and et~al.
\newblock {The Hot and Energetic Universe: A White Paper presenting the science
  theme motivating the Athena+ mission}.
\newblock \emph{ArXiv e-prints}, June 2013.

\bibitem[{Odenwald} et~al.(2003){Odenwald}, {Kashlinsky}, {Mather},
  {Skrutskie}, and {Cutri}]{Odenwald:2003}
S.~{Odenwald}, A.~{Kashlinsky}, J.~C. {Mather}, M.~F. {Skrutskie}, and R.~M.
  {Cutri}.
\newblock {Analysis of the Diffuse Near-Infrared Emission from Two-Micron
  All-Sky Survey Deep Integration Data: Foregrounds versus the Cosmic Infrared
  Background}.
\newblock \emph{\apj}, 583:\penalty0 535--550, February 2003.
\newblock \doi{10.1086/345401}.

\bibitem[{Spergel} et~al.(2015){Spergel}, {Gehrels}, {Baltay}, {Bennett},
  {Breckinridge}, {Donahue}, {Dressler}, {Gaudi}, {Greene}, {Guyon}, {Hirata},
  {Kalirai}, {Kasdin}, {Macintosh}, {Moos}, {Perlmutter}, {Postman},
  {Rauscher}, {Rhodes}, {Wang}, {Weinberg}, {Benford}, {Hudson}, {Jeong},
  {Mellier}, {Traub}, {Yamada}, {Capak}, {Colbert}, {Masters}, {Penny},
  {Savransky}, {Stern}, {Zimmerman}, {Barry}, {Bartusek}, {Carpenter}, {Cheng},
  {Content}, {Dekens}, {Demers}, {Grady}, {Jackson}, {Kuan}, {Kruk}, {Melton},
  {Nemati}, {Parvin}, {Poberezhskiy}, {Peddie}, {Ruffa}, {Wallace}, {Whipple},
  {Wollack}, and {Zhao}]{Spergel:2015}
D.~{Spergel}, N.~{Gehrels}, C.~{Baltay}, D.~{Bennett}, J.~{Breckinridge},
  M.~{Donahue}, A.~{Dressler}, B.~S. {Gaudi}, T.~{Greene}, O.~{Guyon},
  C.~{Hirata}, J.~{Kalirai}, N.~J. {Kasdin}, B.~{Macintosh}, W.~{Moos},
  S.~{Perlmutter}, M.~{Postman}, B.~{Rauscher}, J.~{Rhodes}, Y.~{Wang},
  D.~{Weinberg}, D.~{Benford}, M.~{Hudson}, W.-S. {Jeong}, Y.~{Mellier},
  W.~{Traub}, T.~{Yamada}, P.~{Capak}, J.~{Colbert}, D.~{Masters}, M.~{Penny},
  D.~{Savransky}, D.~{Stern}, N.~{Zimmerman}, R.~{Barry}, L.~{Bartusek},
  K.~{Carpenter}, E.~{Cheng}, D.~{Content}, F.~{Dekens}, R.~{Demers},
  K.~{Grady}, C.~{Jackson}, G.~{Kuan}, J.~{Kruk}, M.~{Melton}, B.~{Nemati},
  B.~{Parvin}, I.~{Poberezhskiy}, C.~{Peddie}, J.~{Ruffa}, J.~K. {Wallace},
  A.~{Whipple}, E.~{Wollack}, and F.~{Zhao}.
\newblock {Wide-Field InfrarRed Survey Telescope-Astrophysics Focused Telescope
  Assets WFIRST-AFTA 2015 Report}.
\newblock \emph{ArXiv e-prints}, March 2015.

\bibitem[{The Simons Observatory Collaboration} et~al.(2018){The Simons
  Observatory Collaboration}, {Ade}, {Aguirre}, {Ahmed}, {Aiola}, {Ali},
  {Alonso}, {Alvarez}, {Arnold}, {Ashton}, and
  et~al.]{The-Simons-Observatory-Collaboration:2018}
{The Simons Observatory Collaboration}, P.~{Ade}, J.~{Aguirre}, Z.~{Ahmed},
  S.~{Aiola}, A.~{Ali}, D.~{Alonso}, M.~A. {Alvarez}, K.~{Arnold}, P.~{Ashton},
  and et~al.
\newblock {The Simons Observatory: Science goals and forecasts}.
\newblock \emph{arXiv e-prints}, August 2018.

\bibitem[{Thompson} et~al.(2007{\natexlab{a}}){Thompson}, {Eisenstein}, {Fan},
  {Rieke}, and {Kennicutt}]{Thompson:2007}
R.~I. {Thompson}, D.~{Eisenstein}, X.~{Fan}, M.~{Rieke}, and R.~C. {Kennicutt}.
\newblock {Evidence for a $z < 8$ Origin of the Source-subtracted Near-Infrared
  Background}.
\newblock \emph{\apj}, 666:\penalty0 658--662, September 2007{\natexlab{a}}.
\newblock \doi{10.1086/520634}.

\bibitem[{Thompson} et~al.(2007{\natexlab{b}}){Thompson}, {Eisenstein}, {Fan},
  {Rieke}, and {Kennicutt}]{Thompson:2007a}
R.~I. {Thompson}, D.~{Eisenstein}, X.~{Fan}, M.~{Rieke}, and R.~C. {Kennicutt}.
\newblock {Constraints on the Cosmic Near-Infrared Background Excess from
  NICMOS Deep Field Observations}.
\newblock \emph{\apj}, 657:\penalty0 669--680, March 2007{\natexlab{b}}.
\newblock \doi{10.1086/511380}.

\bibitem[{Thornton} et~al.(2016){Thornton}, {Ade}, {Aiola}, {Angil{\`e}},
  {Amiri}, {Beall}, {Becker}, {Cho}, {Choi}, {Corlies}, {Coughlin}, {Datta},
  {Devlin}, {Dicker}, {D{\"u}nner}, {Fowler}, {Fox}, {Gallardo}, {Gao},
  {Grace}, {Halpern}, {Hasselfield}, {Henderson}, {Hilton}, {Hincks}, {Ho},
  {Hubmayr}, {Irwin}, {Klein}, {Koopman}, {Li}, {Louis}, {Lungu}, {Maurin},
  {McMahon}, {Munson}, {Naess}, {Nati}, {Newburgh}, {Nibarger}, {Niemack},
  {Niraula}, {Nolta}, {Page}, {Pappas}, {Schillaci}, {Schmitt}, {Sehgal},
  {Sievers}, {Simon}, {Staggs}, {Tucker}, {Uehara}, {van Lanen}, {Ward}, and
  {Wollack}]{Thornton:2016}
R.~J. {Thornton}, P.~A.~R. {Ade}, S.~{Aiola}, F.~E. {Angil{\`e}}, M.~{Amiri},
  J.~A. {Beall}, D.~T. {Becker}, H.-M. {Cho}, S.~K. {Choi}, P.~{Corlies}, K.~P.
  {Coughlin}, R.~{Datta}, M.~J. {Devlin}, S.~R. {Dicker}, R.~{D{\"u}nner},
  J.~W. {Fowler}, A.~E. {Fox}, P.~A. {Gallardo}, J.~{Gao}, E.~{Grace},
  M.~{Halpern}, M.~{Hasselfield}, S.~W. {Henderson}, G.~C. {Hilton}, A.~D.
  {Hincks}, S.~P. {Ho}, J.~{Hubmayr}, K.~D. {Irwin}, J.~{Klein}, B.~{Koopman},
  D.~{Li}, T.~{Louis}, M.~{Lungu}, L.~{Maurin}, J.~{McMahon}, C.~D. {Munson},
  S.~{Naess}, F.~{Nati}, L.~{Newburgh}, J.~{Nibarger}, M.~D. {Niemack},
  P.~{Niraula}, M.~R. {Nolta}, L.~A. {Page}, C.~G. {Pappas}, A.~{Schillaci},
  B.~L. {Schmitt}, N.~{Sehgal}, J.~L. {Sievers}, S.~M. {Simon}, S.~T. {Staggs},
  C.~{Tucker}, M.~{Uehara}, J.~{van Lanen}, J.~T. {Ward}, and E.~J. {Wollack}.
\newblock {The Atacama Cosmology Telescope: The Polarization-sensitive ACTPol
  Instrument}.
\newblock \emph{\apjs}, 227:\penalty0 21, December 2016.
\newblock \doi{10.3847/1538-4365/227/2/21}.

\bibitem[{Ward} et~al.(2016){Ward}, {Austermann}, {Beall}, {Choi}, {Crowley},
  {Devlin}, {Duff}, {Gallardo}, {Henderson}, {Ho}, {Hilton}, {Hubmayr},
  {Khavari}, {Klein}, {Koopman}, {Li}, {McMahon}, {Mumby}, {Nati}, {Niemack},
  {Page}, {Salatino}, {Schillaci}, {Schmitt}, {Simon}, {Staggs}, {Thornton},
  {Ullom}, {Vavagiakis}, and {Wollack}]{Ward:2016}
J.~T. {Ward}, J.~{Austermann}, J.~A. {Beall}, S.~K. {Choi}, K.~T. {Crowley},
  M.~J. {Devlin}, S.~M. {Duff}, P.~A. {Gallardo}, S.~W. {Henderson}, S.-P.~P.
  {Ho}, G.~{Hilton}, J.~{Hubmayr}, N.~{Khavari}, J.~{Klein}, B.~J. {Koopman},
  D.~{Li}, J.~{McMahon}, G.~{Mumby}, F.~{Nati}, M.~D. {Niemack}, L.~A. {Page},
  M.~{Salatino}, A.~{Schillaci}, B.~L. {Schmitt}, S.~M. {Simon}, S.~T.
  {Staggs}, R.~{Thornton}, J.~N. {Ullom}, E.~M. {Vavagiakis}, and E.~J.
  {Wollack}.
\newblock {Mechanical designs and development of TES bolometer detector arrays
  for the Advanced ACTPol experiment}.
\newblock In \emph{Millimeter, Submillimeter, and Far-Infrared Detectors and
  Instrumentation for Astronomy VIII}, volume 9914 of \emph{\procspie}, page
  991437, July 2016.
\newblock \doi{10.1117/12.2233746}.

\bibitem[{Windhorst} et~al.(2018){Windhorst}, {Timmes}, {Wyithe}, {Alpaslan},
  {Andrews}, {Coe}, {Diego}, {Dijkstra}, {Driver}, {Kelly}, and
  {Kim}]{Windhorst:2018}
R.~A. {Windhorst}, F.~X. {Timmes}, J.~S.~B. {Wyithe}, M.~{Alpaslan}, S.~K.
  {Andrews}, D.~{Coe}, J.~M. {Diego}, M.~{Dijkstra}, S.~P. {Driver}, P.~L.
  {Kelly}, and D.~{Kim}.
\newblock {On the Observability of Individual Population III Stars and Their
  Stellar-mass Black Hole Accretion Disks through Cluster Caustic Transits}.
\newblock \emph{\apjs}, 234:\penalty0 41, February 2018.
\newblock \doi{10.3847/1538-4365/aaa760}.

\bibitem[{Yue} et~al.(2013){Yue}, {Ferrara}, {Salvaterra}, {Xu}, and
  {Chen}]{Yue:2013}
B.~{Yue}, A.~{Ferrara}, R.~{Salvaterra}, Y.~{Xu}, and X.~{Chen}.
\newblock {Infrared background signatures of the first black holes}.
\newblock \emph{\mnras}, 433:\penalty0 1556--1566, August 2013.
\newblock \doi{10.1093/mnras/stt826}.

\bibitem[{Yue} et~al.(2014){Yue}, {Ferrara}, {Salvaterra}, {Xu}, and
  {Chen}]{Yue:2014}
B.~{Yue}, A.~{Ferrara}, R.~{Salvaterra}, Y.~{Xu}, and X.~{Chen}.
\newblock {The brief era of direct collapse black hole formation}.
\newblock \emph{\mnras}, 440:\penalty0 1263--1273, May 2014.
\newblock \doi{10.1093/mnras/stu351}.

\bibitem[{Yue} et~al.(2016){Yue}, {Ferrara}, and {Salvaterra}]{Yue:2016a}
B.~{Yue}, A.~{Ferrara}, and R.~{Salvaterra}.
\newblock {Updated analysis of near-infrared background fluctuations}.
\newblock \emph{ArXiv e-prints}, January 2016.

\bibitem[{Zemcov} et~al.(2014){Zemcov}, {Smidt}, {Arai}, {Bock}, {Cooray},
  {Gong}, {Kim}, {Korngut}, {Lam}, {Lee}, {Matsumoto}, {Matsuura}, {Nam},
  {Roudier}, {Tsumura}, and {Wada}]{Zemcov:2014}
M.~{Zemcov}, J.~{Smidt}, T.~{Arai}, J.~{Bock}, A.~{Cooray}, Y.~{Gong}, M.~G.
  {Kim}, P.~{Korngut}, A.~{Lam}, D.~H. {Lee}, T.~{Matsumoto}, S.~{Matsuura},
  U.~W. {Nam}, G.~{Roudier}, K.~{Tsumura}, and T.~{Wada}.
\newblock {On the origin of near-infrared extragalactic background light
  anisotropy}.
\newblock \emph{Science}, 346:\penalty0 732--735, November 2014.
\newblock \doi{10.1126/science.1258168}.

\end{thebibliography}

\end{document}